\documentclass[aps,twocolumn,showpacs,amsmath,amssymb]{revtex4}
\usepackage{dcolumn}
\usepackage{bm}
\usepackage{color}
\usepackage{latexsym}
\usepackage{amssymb}
\usepackage{graphicx}
\usepackage{ulem}

\begin{document}
\title{Magnetoplasmon spectrum for realistic off-plane structure of dissipative 2D system}
\author{M. V. Cheremisin}

\affiliation{A.F.Ioffe Physical-Technical Institute, 194021
St.Petersburg, Russia}

\begin{abstract}
The rigorous analysis of textbook result(Chiu and Quinn, 1974) gives unexpectedly the dramatic change of
magnetoplasmon spectrum taking into account both the arbitrary
dissipation and asymmetric off-plane structure of 2D system. For given wave vector
the dissipation enhancement leads to decrease(increase) of magnetoplasmon frequency
at low(high) magnetic field. At certain range of disorder the purely relaxational mode appears in magnetoplasmon spectrum.
In strong magnetic fields the magnetoplasmon frequency falls to cyclotron resonance line
even in presence of finite dissipation. The recent observation of
2D magnetoplasmon spectrum is consistent with our findings.
\end{abstract}


\maketitle
\section{Introduction}
\label{Introduction}
Plasma oscillations in two-dimensional electron gas(2DEG) were
first predicted in the middle 60th \cite{Stern67},\cite{Chaplik72},\cite{Chiu74} and, then observed
experimentally in liquid helium system \cite{Grimes76} and silicon
inversion layers \cite{Allen77},\cite{Theis78}. The observation
\cite{Kukushkin03} of the magnetoplasmon spectrum reported to be affected
by retardation effects\cite{Chiu74} recommences the interest to the above problem. It was argued that in
large-mesa 2D systems the role of edges becomes less significant, therefore
the observed MP features can be accounted\cite{Cheremisin04} within
conventional theory\cite{Chiu74} for unbounded 2DEG. However, the simple model
of the effective dielectric function fails to account for a peculiar behavior
of the magnetoplasmon spectrum. In present paper we provide the rigorous analysis of
the magnetoplasmon spectrum taking into account both the realistic off-plane asymmetry and,
moreover, arbitrary dissipation of 2DEG. Our results are in a good agrement
with experimental findings\cite{Kukushkin03}.

\section{2D plasmon dispersion law influenced by dissipation and dielectric permittivity mismatch}
\label{Chapter1}
\begin{figure}[tbp]
\begin{center}
\includegraphics[scale=0.5]{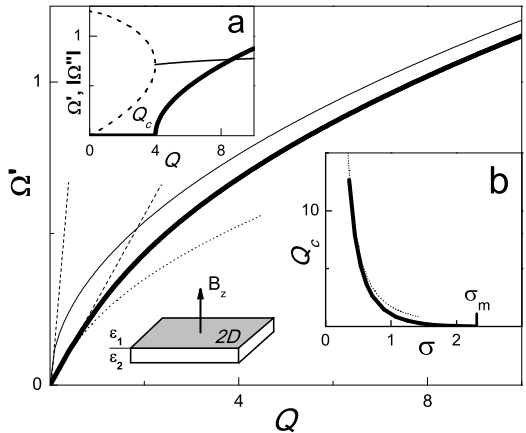}
\caption[]{\label{Fig1} Main panel: The longitudinal plasmon dispersion(bold line) for lossless GaAs-based 2D structure
($\epsilon_{1}=1$,$\epsilon_{2}=12.8$). Left(right)-hand dashed
line depicts the light dispersion $\Omega=Q/\sqrt{\epsilon_{1}}$ and $\Omega=Q/\sqrt{\epsilon_{2}}$ respectively. Thin line
shows high-$Q$ asymptote $\Omega=\sqrt{\frac{2Q}{\epsilon_{1}+\epsilon_{2}}}$. Dotted line corresponds to symmetric 2D structure $\epsilon_{1,2}=12.8$ Inset (a): bold(thin) line depicts the real $\Omega'$ (imaginary $|\Omega''|$) part of the plasmon
frequency for disorder strength $\sigma=\sigma_{c}=0.59$( corresponds to $Q_{c}=4$). Dashed line represents the damping of
relaxational plasmon when $\Omega'=0$. Inset (b): critical wave vector $Q_{c}$ vs dissipation parameter $\sigma< \sigma_{m}=2.3$.
Dotted line corresponds to high-$Q$ asymptote. The art picture represents the asymmetric 2D structure.}
\end{center}
\end{figure}

Let us assume 2DEG in x-y plane(see Fig.\ref{Fig1}, inset) embedded into the dielectric media
with the permittivities $\epsilon_{1}$ and $\epsilon_{2}$ of 1,2-halfspace respectively.
In perpendicular magnetic field, $B\|z$, the complete set of Maxwell
equations for in-plane components of the electrodynamic potentials $\bf{A}, \phi$
yields \cite{Falko89}
\begin{eqnarray}
\fbox{}\phi=4\pi\rho, \fbox{}{\bf A}=\frac{4\pi {\bf j}}{c},
\label{Maxwell} \\
\text{div}{\bf A}+\frac{\epsilon}{c}\frac{\partial \phi}{\partial t}=0, \nonumber \\
{\bf j}=-\hat{\sigma}\left( \nabla \phi +\frac{1}{c}\frac{\partial
{\bf A}}{\partial t} \right), \nonumber
\end{eqnarray}
where $\fbox{}=\frac{\epsilon}{c^{2}}\frac{\partial^{2}}{\partial
t^{2}}-\Delta$ is the d'Alambert operator. In presence of the magnetic field the conductivity
tensor $\hat{\sigma}$ contains the longitudinal $\sigma_{xx}=\sigma_{yy}$ and transverse
$\sigma_{yx}=-\sigma_{xy}$ components.

Assuming the magnetoplasmon $e^{(i{\bf qr}-i\omega t)}$ propagated in 2DEG, and, then separating
the longitudinal and the transverse in-plane components of the vector
potential\cite{Falko89}, the magnetoplasmon dispersion relation yields:
\begin{equation}
\left[ \frac{1}{4\pi}\left ( \frac{\epsilon_{1}}{\kappa_{1}}+\frac{\epsilon_{2}}{\kappa_{2}} \right ) +\frac{i \sigma_{xx} }{\omega}
\right ]
\left[\frac{\kappa_{1}+\kappa_{2}}{4\pi}-\frac{i \omega \sigma_{xx}}{c^{2}} \right ]+ \frac{\sigma_{yx}^{2}}{c^{2}}=0, \\
\label{MP_dispersion}
\end{equation}
where $\kappa_{1,2}=\sqrt{q^{2}-\epsilon_{1,2} \frac{\omega^{2}}{c^{2}}}>0$ denotes the inverse penetration
length of electromagnetic fields $\sim e^{-\kappa |z|}$ into 1(2)-halfspace respectively. Note that Eq.(\ref{MP_dispersion})
was first derived by Chiu and Quinn\cite{Chiu74}. Until now, Eq.(\ref{MP_dispersion}) was analyzed in shortcut form
for off-plane symmetric 2DEG sample.

Let us specify the components of the conductivity tensor embedded into Eq.(\ref{MP_dispersion}).
Following conventional Drude formalism one can represent them as it follows
\begin{equation}
\sigma_{xx}=\frac{i\tilde{\Omega}}{\tilde{\Omega}^{2}-\Omega^{2}_{c}}\frac{\sigma_{0}}{\sigma},
\quad \sigma_{yx}=\frac{i\Omega_{c}}{\tilde{\Omega}}\sigma_{xx},
\label{conductivity}\\
\end{equation}
where $\sigma_{0}=ne^{2}\tau/m$ is the Drude conductivity at $B=0$, $n$ is the density, $m$ is the effective mass,
and $\tau$ is the momentum relaxation time. Then, for actual problem of magnetoplasmon spectrum we use the notations made of use in Ref.\cite{Cheremisin04}.
Namely, we specify the dimensionless frequency $\Omega=\frac{\omega}{\omega_{p}}$ and the cyclotron frequency $\Omega_{c}=\frac{\omega_{c}}{\omega_{p}}$
scaled by the dimensional unit $\omega_{p}=\frac{2\pi ne^{2}}{mc}$. In addition, we use the auxiliary quantity $\tilde{\Omega}=\Omega+i/\sigma$,
where the dimensionless dissipation parameter $\sigma=\frac{2\pi\sigma_{0}}{c}$ is known to be a measure of 2DEG charge relaxation dynamics
in presence of retardation effects\cite{Dyakonov87,Govorov89}.

In absence of the magnetic field, i.e. when $\sigma_{yx}=0$, Eq.(\ref{MP_dispersion}) decouples. The left(right)-hand term in square brackets
defines the dispersion law for longitudinal (transverse) plasmon respectively. The transverse mode is purely relaxational\cite{Falko89}. Hence,
our primary interest concerns the dispersion law
\begin{equation}
\frac{\epsilon_{1}}{\kappa_{1}}+\frac{\epsilon_{2}}{\kappa_{2}}=\frac{2}{\Omega \tilde{\Omega}}
\label{plasmonB0}
\end{equation}
for longitudinal mode which could demonstrate the weakly damped or purely relaxational behavior dependent on dissipation strength.
Here, $\kappa_{1,2}=\sqrt{Q^{2}-\epsilon_{1,2}\Omega^{2}}$ is the dimensionless inverse penetration length,
$Q=\frac{qc}{\omega_{p}}$ is the dimensionless wave vector.

We emphasize that the plasmon dispersion law specified by Eq.(\ref{plasmonB0}) is affected by the dielectric mismatch
$\epsilon_{1}\neq \epsilon_{2}$. To confirm this, we solve numerically Eq.(\ref{plasmonB0})
for typical GaAs-based 2DEG\cite{Kukushkin03} which exhibits strong asymmetry, i.e. $\epsilon_{1}=1$,
$\epsilon_{2}=12.8$. Let us first consider the dissipationless carriers case, when $\sigma \rightarrow \infty $. The result
is represented by the bold line in Fig.\ref{Fig1}, main panel. At high values of the wave vector $Q \gg \sqrt{\epsilon_{1,2}} \Omega$
the retardation effects can be ignored. The plasmon spectrum(see thin line in Fig.\ref{Fig1}, main panel)
obeys the familiar square-root dispersion relationship $\Omega=\sqrt{\frac{2Q}{\epsilon_{1}+\epsilon_{2}}}$ with the average permittivity embedded.
This intuitive result is known in literature as "effective dielectric function approach" and claimed to be universal.
We argue, however, that in the opposite low-$Q$ case the average permittivity scenario fails to account the plasmon spectrum.
Indeed, at $Q \leq \sqrt{\epsilon_{1,2}} \Omega$ the retardation effects becomes of the primary importance. Consequently, the plasmon
dispersion curve in Fig.\ref{Fig1} is located well below the lowest light dispersion asymptote $\Omega=Q/\sqrt{\epsilon_{2}}$
associated with GaAs bulk of actual 2D structure. We argue that the common use of the "effective dielectric function approach" is well justified
for high-$Q$ part of the plasmon spectrum only. In opposite low-$Q$ case the retardation effects become highly important, therefore
the problem of electromagnetic fields in the vicinity of 2DEG plane needs to be solved exactly.

We now examine the plasmon spectrum for real case of dissipative 2D system. At fixed disorder strength
the solution of Eq.(\ref{plasmonB0}) gives of the complex frequency $\Omega=\Omega'+i\Omega''$. The plasmon is
damped when $\Omega''<0$. We verify that for arbitrary dissipation the plasmon electromagnetic fields
are indeed localized within 2D plane since $\mathrm{Re}(\kappa_{1,2})>0$.

In Fig.\ref{Fig1},a we plot both the real and imaginary part of the longitudinal plasmon frequency vs wave vector
for certain value of the dissipation strength. The evidence shows that the real component of the frequency becomes
positive when $Q>Q_{c}$, where $Q_{c}$ is a critical wave vector which, in turn, can be represented
as a function of $\sigma$ and $\epsilon_{1,2}$. For actual GaAs-based 2DEG
we find numerically the critical diagram $Q_{c}(\sigma)$ shown in Fig.\ref{Fig1},b.
The part of the diagram above the bold line corresponds to plasmon excitations, i.e. when $\Omega'>0$. Note that for certain
wave vector $Q$ the graphic solution of the equation $Q_{c}(\sigma_{c})=Q$ defines a critical value of
dissipation, $\sigma_{c}$, above which the plasmon excitations exist. To confirm this finding we
solve Eq.(\ref{plasmonB0}) and, then plot in Fig.\ref{Fig2}c,d both the real and imaginary part of the longitudinal plasmon frequency vs
disorder parameter assuming fixed wave vector $Q=4$. We find that the longitudinal plasmon exists when $\sigma>\sigma_{c}=0.59$.
For stronger disorder $\sigma<\sigma_{c}$ the only relaxational mode exists. Substituting $\Omega'=0$ into Eq.(\ref{plasmonB0})
one can easily obtain the longitudinal plasmon damping $\Omega''(\sigma)$ shown by dashed line in Fig.\ref{Fig2}d.
Similarly, we add in Fig.\ref{Fig2}d the transverse plasmon damping which follows from the right-hand square brackets term
in Eq.(\ref{MP_dispersion}). We conclude that for high disorder case $\sigma<\sigma_{c}$ there exists a three
different relaxational modes displayed, for example, by open symbols in Fig.\ref{Fig2}d.

Remarkably, the critical diagram $Q_{c}(\sigma)$ in Fig.\ref{Fig1},b can be found analytically within high-$Q$ non-retarded limit.
Indeed, with the help of Eq.(\ref{plasmonB0}) both the real and the imaginary parts of the complex frequency yield
\begin{equation}
\Omega'=\sqrt{\frac{2Q}{\epsilon_{1}+\epsilon_{2}}-\frac{1}{4\sigma^{2}}},\quad \Omega''=-\frac{1}{2\sigma}.
\label{Assympt_High_Q}
\end{equation}
Actually, the Eq.(\ref{Assympt_High_Q}) provides the critical diagram asymptote $Q_{c}(\sigma)=\frac{\epsilon_{1}+\epsilon_{2}}{8\sigma^{2}}$
shown by dotted line in Fig.\ref{Fig1},b.

We emphasize that the critical diagram $Q_{c}(\sigma)$ demonstrates
vanishing at certain magnitude of the dissipation $\sigma_{m}$. Fortunately, the above value can be easily found
within analytic approach as well. Indeed, the substitution $Q,\Omega'=0$ and, moreover, $\Omega''\rightarrow 0$ into Eq.(\ref{plasmonB0})
gives the result $\sigma_{m}=\frac{\sqrt{\epsilon_{1}}+\sqrt{\epsilon_{2}}}{2}$. We conclude that the long-wave plasmon $Q \rightarrow 0$ always
exists when $\sigma>\sigma_{m}$. Note that in particular case of symmetric 2DEG, i.e. when $\epsilon_{1,2}=\epsilon$, our
finding coincides with that $\sigma_{m}=\sqrt{\epsilon}$ reported in Ref.\cite{Falko89}. We argue also that the dimensionless
disorder parameter $\tilde{\sigma}$ made of use in Ref.\cite{Volkov16} for symmetric case is related to that in our
notations via the relationship $\tilde{\sigma}=\sigma/\sqrt{\epsilon}$. As expected, the long-wave plasmon exists when
$\tilde{\sigma}>1$ in notations of Ref.\cite{Volkov16}.

\section{Dramatic change of the magnetoplasmon spectrum caused by off-plane dielectric asymmetry and dissipation}
\label{Chapter2}
The up-to-date attempts to analyze the plasmon spectrum in presence of the magnetic field concern either non-dissipative
\cite{Chiu74} or dissipative 2D plasma\cite{Volkov16}. In both cases the dielectric mismatch of 2D structure was neglected.
We now demonstrate that both the off-plane dielectric asymmetry and the dissipation strongly affect the magnetoplasmon spectrum.
In contrast to authors of Ref.\cite{Volkov16} we intend to consider the regular experimental setup implying the changeless 2D sample geometry.
Consequently, we will analyze the magnetoplasmon excitations for fixed wave vector and varied magnetic field
and compare our results with existing experimental data\cite{Kukushkin03}.

Substituting the components of the conductivity tensor given by Eq.(\ref{conductivity}) into Eq.(\ref{MP_dispersion})
the magnetoplasmon spectrum yields
\begin{equation}
\Omega_{c}=\tilde{\Omega} \left [ \frac{ \left (   \frac{\epsilon_{1}}{\kappa_{1}}+\frac{\epsilon_{2}}{\kappa_{2}}-\frac{2}{\Omega \tilde{\Omega}}  \right )\left ( \kappa_{1}+\kappa_{2}+\frac{2\Omega}{\tilde{\Omega}} \right )}{\left ( \frac{\epsilon_{1}}{\kappa_{1}}+\frac{\epsilon_{2}}{\kappa_{2}} \right )(\kappa_{1}+\kappa_{2})} \right ]^{1/2}.
\label{MP_spectrum_general}
\end{equation}
Evidently, the zero-field dispersion law specified by Eq.(\ref{plasmonB0}) follows from Eq.(\ref{MP_spectrum_general})
when $\Omega_{c}=0$. We argue that for fixed values of dielectric permittivities $\epsilon_{1,2}$, wave vector $Q$ and the dissipation
strength $\sigma$ one can solve Eq.(\ref{MP_spectrum_general}) and, then find both the real and the imaginary
parts of the magnetoplasmon frequency as a functions of the cyclotron one.

\begin{figure}[tbp]
\begin{center}
\includegraphics[scale=0.5]{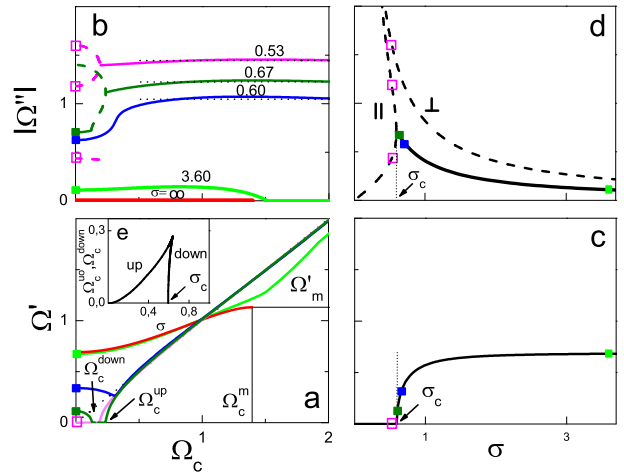}
\caption[]{\label{Fig2} (Color online) Panel a(b) shows the real(imaginary) part of the magnetoplasmon frequency vs
cyclotron frequency at fixed wave vector $Q=4$ for dissipationless $\sigma =\infty$ and lossy $\sigma=3.6;0.67;0.60;0.53$
GaAs-based 2D system($\epsilon_{1}=1$,$\epsilon_{2}=12.8$). The relaxational magnetoplasmon mode(i.e. when $\Omega'=0$) is
represented by dashed lines. Dotted lines depict the high-field
asymptotes $\Omega_{\infty}$ specified by Eq.(\ref{Assympt_High_B}). For lossless 2DEG the spectrum cutoff occurs at
$\Omega'_{m}=1.12$ and $\Omega^{m}_{c}=1.38$.
Solid lines in panel c(d) depict the real(imaginary) part of the frequency for zero-field longitudinal plasmon vs
dissipation strength. Dashed lines are related to relaxational mode for longitudinal($\parallel$) and
transverse($\perp$) plasmon respectively. Square symbols in panel c(d)
correspond to those in the panel a(b). Inset (e): Threshold diagram for relaxational plasma mode.}
\end{center}
\end{figure}

Let us examine first the dissipationless case when magnetoplasmon is undamped. Substitution $\Omega''=0$
into Eq.(\ref{MP_spectrum_general}) readily gives the sought-for spectrum in the form $\Omega_{c}(\Omega',Q)$. For actual
GaAs-based 2DEG the result is shown in Fig.\ref{Fig2}a for fixed wave vector $Q=4$. Remarkably, the lossless
magnetoplasmon excitation exhibits a certain cutoff point $\Omega'_{m},\Omega^{m}_{c}$ on the spectrum plot. Indeed, for asymmetric 2D structure in question
the magnetoplasmon penetration length becomes infinitely large for 2-halfspace when $\kappa_{2}=0$.
Under this condition the Eq.(\ref{MP_spectrum_general}) provides the spectrum cutoff point as $\Omega'_{m}=\frac{Q}{\sqrt{\epsilon_{2}}}$ and
$\Omega^{m}_{c}=\left [\frac{Q^{2}}{\epsilon_{2}}+\frac{2Q}{\sqrt{\epsilon_{2}-\epsilon_{1}}} \right ]^{1/2}$
respectively. Note that for symmetric 2D structure the lossless magnetoplasmon spectrum could demonstrate the saturation\cite{Chiu74,Cheremisin04} at high
magnetic fields since $\Omega^{m}_{c}\rightarrow \infty$.

We now analyze the magnetoplasmon spectrum for fixed wave vector and, moreover, assume a finite disorder.
In general, Eq.(\ref{MP_spectrum_general}) can be solved numerically. The solid curves in Figs.\ref{Fig2}a,b
correspond to magnetoplasmon excitations possessing the nonzero frequency. At low magnetic fields the
real(imaginary) component of the complex frequency decreases(increases)
under the disorder enhancement. This finding correlates with that discussed above for zero-field plasmon at
$\sigma\geq\sigma_{c}$(see Figs.\ref{Fig2}c,d). For given dissipation strength the zero-field plasmon
frequency(damping) shown by the solid symbols in Fig.\ref{Fig2}, panel c(d) corresponds
to those in Figs.\ref{Fig2}, panel a(b) respectively. In contrast, for high-disorder case $\sigma\leq\sigma_{c}$ the only relaxational plasma
modes could appear in 2D system. As an example, the relaxational plasma modes associated
with longitudinal and(or) transverse plasmon are represented by open symbols in Figs.\ref{Fig2}b,d.

We argue that the substantial drop of the lossy magnetoplasmon frequency
at low magnetic fields compared to that expected for clean 2DEG can be attributed to so-called "low-frequency mode"
discussed in literature\cite{Muravev15,Gusikhin14}. Let us analyze in details the behavior of lossy magnetoplasmon spectrum
at moderate magnetic fields. Figs.\ref{Fig2}a,b provides a strong evidence of relaxational mode appeared within a
certain range of magnetic fields $\Omega_{c}^{down}<\Omega_{c}<\Omega_{c}^{up}$ even for $\sigma>\sigma_{c}$.
We find numerically and, then plot in Fig.\ref{Fig2}e the dependencies $\Omega_{c}^{down},\Omega_{c}^{up}$ vs $\sigma$, thus obtain the
relaxational mode diagram. The area outside the triangle in Fig.\ref{Fig2}e corresponds to magnetoplasmon excitations possessing nonzero frequency.
Note that for strong disorder $\sigma<<\sigma_{c}$ the magnetoplasmon frequency approaches the cyclotron line which is seen in Fig.\ref{Fig2}a.

In high magnetic fields the behavior of the magnetoplasmon spectrum is striking as well.
At first, the all curves $\Omega'(\Omega_{c})$ plotted for different dissipation strengths demonstrate a certain crossing with
cyclotron resonance line. Well above the crossing point the magnetoplasmon spectrum exhibits the change from cutoff
trend owned to dissipationless case to linear in magnetic field behavior for lossy magnetoplasmon.
As an example, for simplest case of lossless 2DEG we substitute $\Omega_{c}=\Omega'$ into Eq.(\ref{MP_spectrum_general}) and,
then find the transcendental equation
\begin{equation}
\frac{\epsilon_{1}}{\kappa_{1}}+\frac{\epsilon_{2}}{\kappa_{2}}=\frac{\kappa_{1}+\kappa_{2}+2}{\Omega^{2}}.
\label{Crossing_point}
\end{equation}
for sought-for crossing point frequency. For GaAs 2DEG case shown in Fig.\ref{Fig2}a we obtain the crossing
point at $\Omega'=\Omega_{c}=1.02$.

\begin{figure}[tbp]
\begin{center}
\includegraphics[scale=0.5]{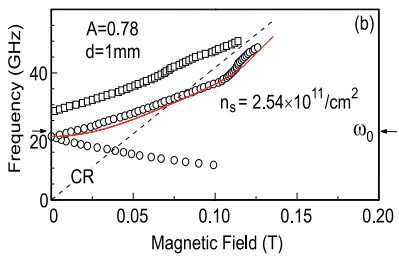}
\caption[]{\label{Fig3} (Color online) Bulk magnetoplasmon spectrum data under Ref.\cite{Kukushkin03} for disk-shaped GaAs 2D sample. Thin line
depicts the numerical result for lowest mode of 2D resonator cavity at $Q=3.6$ and $\sigma=2.8$.}
\end{center}
\end{figure}

We argue that in strong magnetic fields the magnetoplasmon spectrum can exhibit(see Fig.\ref{Fig2}a) a non-monotonic behavior
at moderate dissipation $\sigma \sim 1$. To test our model,
we analyze in Fig.\ref{Fig3}  the experimental data\cite{Kukushkin03}. For actual carrier density $n=2.54\times 10^{11}$cm$^{-2}$ and
mesa diameter $d=1$mm of GaAs 2D sample, we find the frequency $\omega_{p}=2 \times 10^{11}$c$^{-1}$
and the wave vector $q=2.4/d=24$cm$^{-1}$, thus $Q=3.6$. The best fit of the lowest 2D resonator cavity mode is shown in Fig.\ref{Fig3}
for disorder strength $\sigma=2.8$. The later allows us to estimate the carrier mobility $\sim 0.37\times 10^{6}$cm$^{2}$/Vs
being close to that $\sim 10^{6}$cm$^{2}$/Vs reported in experiment\cite{Kukushkin03}.

Remarkably, for finite dissipation strength we are able to solve analytically Eq.(\ref{MP_spectrum_general})
within high frequency limit $\Omega\gg Q/\sqrt{\epsilon_{1,2}}$. Indeed, one can represent the inverse penetration
length as $\kappa_{1,2}=i\sqrt{\epsilon_{1,2}}\Omega \left (1-\frac{Q^{2}}{\Omega^{2}\epsilon_{1,2}}\right )^{1/2}$.
Keeping $Q^{4}$-order terms the complex frequency yields
\begin{eqnarray}
\Omega_{\infty}=\Omega_{c}+i(\sigma_{m}^{-1}-\sigma^{-1}),
\nonumber \\
\Omega=\Omega_{\infty}+\frac{Q^{4}(i\Omega_{c}-\sigma_{m}^{-1})}{16\sigma_{m}^{2}\Omega_{\infty}^{4}\Omega_{c}}
\left (\epsilon_{1}^{-3/2}+\epsilon_{2}^{-3/2} \right ).
\label{Assympt_High_B}
\end{eqnarray}
Eq.(\ref{Assympt_High_B}) is valid when $\sigma<\sigma_{m}$ whereas $\Omega_{\infty}$ denotes the magnetoplasmon frequency
at $B\rightarrow \infty$. Note that for symmetric 2DEG Eq.(\ref{Assympt_High_B}) coincides with Eq.(3) obtained in Ref.\cite{Volkov16}.

According to Eq.(\ref{Assympt_High_B}) at high magnetic fields the magnetoplasmon frequency follows the cyclotron resonance asymptote,
while the damping depends on dissipation strength. To confirm this finding, in Fig.\ref{Fig2} we plot the asymptotes
specified by Eq.(\ref{Assympt_High_B}). Numerical data is well described by theory predictions.
We emphasize also that at high magnetic fields the magnetoplasmon is always localized nearby 2D plane.
Indeed, Eq.(\ref{Assympt_High_B}) allows one to find the correct inverse penetration length of the electromagnetic
fields for both sides of 2D plane, i.e. $\mathrm{Re}(\kappa_{1,2})\simeq \sqrt{\epsilon_{1,2}}|\Omega''|>0$.

\section{Conclusions}
\label{Conclusions}
Our analysis of textbook Eq.(\ref{MP_dispersion}) derived in early 70-s\cite{Chiu74} provides
the strong doubts concerning overall use of effective dielectric function approach. We demonstrate
the dramatic change of magnetoplasmon spectrum for realistic case of asymmetric off-plane structure of dissipative
2D systems. Under disorder enhancement the magnetoplasmon with a certain wave vector demonstrates the
decrease (increase) of its frequency at low (high) magnetic field. At certain range of disorder the
purely relaxational mode appears in magnetoplasmon spectrum. At high magnetic fields the real(imaginary) component
of magnetoplasmon frequency follows the cyclotron resonance line and depends on the dissipation strength
respectively. For moderate dissipation our calculations provide an evidence of non-monotonic behavior
of magnetoplasmon spectrum already observed in experiment.

\end{document}